# BEAM LOSS CONTROL FOR THE FERMILAB MAIN INJECTOR*

Bruce C. Brown[†], Fermilab, Batavia, IL 60510, USA

*Abstract*

From 2005 through 2012, the Fermilab Main Injector provided intense beams of 120 GeV protons to produce neutrino beams and antiprotons. Hardware improvements in conjunction with improved diagnostics allowed the system to reach sustained operation at 400 kW beam power. Losses were at or near the 8 GeV injection energy where 95% beam transmission results in about 1.5 kW of beam loss. By minimizing and localizing loss, residual radiation levels fell while beam power was doubled. Lost beam was directed to either the collimation system or to the beam abort. Critical apertures were increased while improved instrumentation allowed optimal use of available apertures. We will summarize the impact of various loss control tools and the status and trends in residual radiation in the Main Injector.

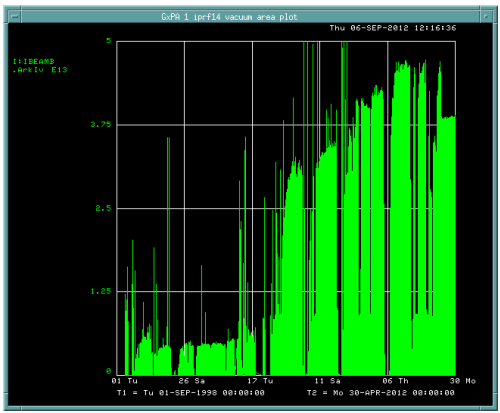

Figure 1: Sampled Intensity per cycle from September 1998 through April 2012.

## PROTONS TO PRODUCE NEUTRINOS AND ANTIPROTONS

On April 30, 2012, the Fermilab accelerator complex began an extended shutdown. This followed seven months after the end of operation for the Tevatron on September 30, 2011 with the accompanying end of antiproton source operation. For the Fermilab Main Injector, this marked $11\frac{1}{2}$ years of commissioning and operation in successively higher intensity operation modes. As the physics program requirements demanded more beam power, limitations in the intensity and beam quality from the Fermilab Booster were overcome by using slip stacking injection[1]. This was implemented first for antiproton production and later for neutrino production as well. As intensities increased, a program of monitoring and mitigating losses and residual radiation has controlled the radiation exposure for personnel involved in maintenance and upgrade activities.

Figure 1 illustrates this intensity increase using the number of protons per cycle on a periodic sample of the acceleration cycles. An injection from the Booster is termed a 'batch' with typical intensity of $4$–$5 \times 10^{12}$ protons and up to 84 rf buckets of beam. Machine commissioning was followed by multibatch operation for a Tevatron fixed target run. In 2001, this transitioned to a Tevatron collider run which utilized a single batch from the Booster for pbar production. Slip stacking injection of two Booster batches for pbar production was developed in 2004. Injecting two batches into buckets of different frequency allows momentum stacking when the buckets slip into alignment and the beam is recaptured in a larger rf bucket. The NuMI beamline for neutrino production was commissioned in 2005 requiring acceleration in each cycle of 5 Booster batches for NuMI in addition to a double batch for pbar production on each cycle (5 plus 2). Slip stacking for increased NuMI beam (9 plus 2) was commissioned in 2007 as was the Main Injector collimation system. At that point intensity was limited by losses in both the Main Injector and the Booster. Collimation, along with improved Booster beam quality, controlled activation and permitted Main Injector intensity per cycle to increase.

Several other features of the Fermilab HEP program are apparent in Fig. 1. Facility upgrades are accomplished using shutdown periods of several weeks. Periods of reduced intensity mark the times required to repair or replace the NuMI horn or target. When pbar production was ended neutrino target intensity limits due to thermal shock could be met by accelerating 9 batches with only three being slip stacked. Reduced per pulse intensity from September 2011 through April 2012 reflects this limitation. The spikes which report exceptionally higher intensity are instrumental. The data uses some of the instrumentation which was replaced by 2007 (see below) and spike above the trend are typically due to instrumentation or data recording errors.

Preparations for the high intensity operation for neutrino production included a program to identify residual radiation issues in the Main Injector tunnel. Exploratory residual radiation measurements in 2004 and 2005 monitored more than 100 locations with more than 20 milliRad/hr residual radiation on contact. By October 2005, a program using a sensitive meter to monitor 127 (later 142) bar-coded locations was initiated[2]. By October 2006, new electronic readout for the beam loss monitors was available[3].

---

Operation console programs for making this data available have been used to display the loss data, allowing improved loss control[4]. Aperture improvements along with collimation systems for loss localization were added in 2006 and 2007. As details of the loss mechanisms were understood, beam manipulations including anti-damping and gap-clearing were added to allow further loss reductions. These loss control tools will be reviewed in turn to provide an overview of loss control for the Main Injector.

## INSTRUMENTATION

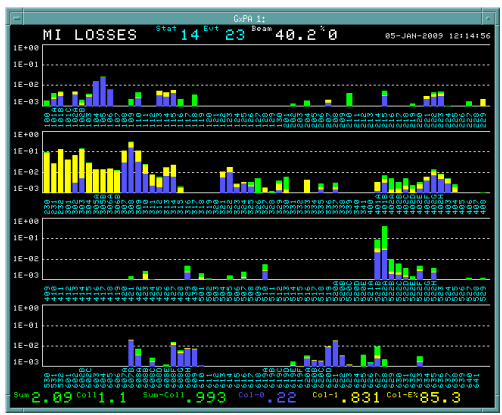

Figure 2: Beam Loss Monitor Display from Main Injector Cycle in January 2009 at an intensity of $4 \times 10^{13}$ protons. Blue shows loss before acceleration, yellow after 1% acceleration, green at end of cycle. Losses in left of upper row are from injection kicker gap. Left end of second row shows collimation loss. Loss at right end of third row are at 522 extraction Lambertson; loss at left of fourth row are at 608 extraction Lambertson; loss at right of fourth row are at 620 Lambertson.

The Main Injector was commissioned using data acquisition systems developed for the Fermilab Main Ring in the 1980's. New instrumentation was commissioned in 2006 for beam loss monitors (BLM's)[3] and beam position monitors (BPM's)[5]. In addition, a more sophisticated data collection system from the existing beam current monitors was developed using a stand alone microprocessor (BEAMS front-end[6]). Together these new instruments allowed a more systematic study of the machine and improved displays of routine operation. Beam loss displays were particularly significant for improving the overall loss pattern by disclosing lesser beam loss locations which had previously gone unobserved. Figure 2 shows beam losses for operations in January 2009. This display occupied a prominent place in the Fermilab Accelerator Control Room. Many losses were reduced by only applying the orbit correction system.

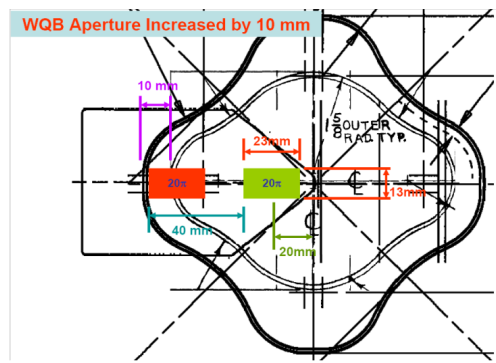

Figure 3: Aperture Improvement using WQB large aperture quadrupoles.

## APERTURE IMPROVEMENTS

### Wide Aperture Quadrupoles

Unsurprisingly, improved apertures allow reduced losses. A major upgrade of the Main Injector apertures was accomplished in 2006 by installing a few quadrupoles with much improved apertures at beam transfer locations. The Lambertson-style septum magnets at injection, abort and extraction points place a septum near the center of nearby quadrupole apertures. Large aperture quadrupole magnets [7, 8] permit the circulating beam orbit to be placed in a region much further from the beam pipe but also in a region of much better magnetic field quality. Figure 3 shows the aperture now provided in one of these transfer locations. The new beam pipe is illustrated by the 'star-shaped' pipe surrounding other features. Beam apertures through the Lambertson magnets are left (circulating) and right (transferred) of center. The beam pipe for the standard quadrupole which were used previously in transfer regions is the smaller star shaped pipe. The injected beam size is shown with the new range of available positions.

### Beam Pipe Alignment

More surprising to us was the discovery of beam pipe misalignment issues. Design vertical apertures at high beta appeared to have several millimeters of clearance from the beam at the three sigma beam boundary. Although a nominal alignment tolerance of 0.25 mm was applied to magnetic devices, it was expected that beam pipe placement would be adequate with only routine placement at support points. As we explored an unexplained loss downstream of the abort Lambertson magnets, we discovered misalignments up to 6 mm. Beam steering following the proper placement of these beam pipes greatly reduced the loss in this area.

A more subtle problem was the first result of the radiation monitoring effort. It was noted that an distinctive loss pattern was evident at more than 10% of the vertically focusing regular half cells. It was characterized by a very local ( 2 cm diameter) high residual radiation on the top of the beam pipe about 0.5 m downstream of the upstream dipole. Analysis of this problem revealed a pair of small mechanical effects combined to make the aperture reduction sufficient for a pattern to emerge[9]. The Main Injector vacuum pipe flexes under atmospheric pressure load, reducing the vertical size by nearly 3 mm. Upstream of each regular cell quadrupole, it is supported at the end of the dipole by a bellows assembly. From that support point, vacuum pipe begins flexing under vacuum tapering to its minimum size at a point 0.5 m along the beam direction. Additionally, the elliptical beam pipe in the troubled locations had been installed through the pre-existing beam pipe of the quadrupole. For reasons not fully understood, some beam pipes were stressed in this process and preferentially stress relieved by bending down. Mis-alignment of these short beam pipes by up to 3 mm combined with the 1.5 mm effect of flexing under vacuum produced the characteristic localized radiation pattern.

*Instrumentation Problems*

The improved sensitivity provided by new BPM's facilitated some studies which required the better resolution. In doing these measurements, we discovered a BPM detector which had faulty connections. This error had resulted in setting the orbit to wrong locations by up to 15 mm. The large horizontal aperture of the Main Injector allowed adequate transmission despite such errors but losses improved when this was corrected. Occasional other BPM failures were quickly noted after implementation of the beam loss display. Occasional BLM failures also allowed some additional activation before they were located.

## COLLIMATION

*Injection Line Collimation*

In order to collimate beam in a transfer line, one requires collimation edges on two sides of the beam and at two locations. This was accomplished in the Fermilab Booster to Main Injector transfer line (MI8 Line) using the corners of four rectangular apertures separated by $90^o$ phase advance. This collimation system[10] was installed in 2006 and has operated to scrape beam edges beyond about 99% of the beam. Beam motion would cause fluctuations in the transmitted beam by asymmetric collimation. This was controlled by an auto-tune system using beam position and trim magnet settings with rapid updates resulting in stability at the $\sim 0.1$ mm level.

*Main Injector Ring Collimation*

Based on the successes of the transfer line collimators, a system with similar mechanical properties was installed in the MI300 section of the Main Injector[11, 12]. It was designed using a primary-secondary collimation scheme such that the unaccelerated beam which was outside of the accelerating bucket after slip stacking high voltage recapture would strike the 0.25 mm tungsten primary collimator which sits at the last cell with high dispersion ahead of a long straight section. The scattered beam strikes one of four secondary collimators which are placed at appropriate phase advance downstream. This loss pattern is distinctive due to the narrow time structure of the 8 GeV beam moving to the low momentum dispersion orbit. Using the time structure as a diagnostic, examination of the ring loss pattern shows that 99% of the radiation from this beam loss is captured in the collimation region[11]. With the longitudinal emittance of the Booster at $4.3 \times 10^{12}$ protons per Booster cycle, this loss is about 5% of the injected beam.

## GAP CLEARING

Some beam from the slip stacking process moves from the design buckets to regions where beam gaps are placed to permit transfers with fast magnets (kickers) for extraction or injection. Unless the problem is mitigated, the beam in injection gaps will be driven into magnets (MI104 through MI106) near the injection kickers (MI103). Beam in extraction gaps would strike the Lambertson magnets near the MI522 transfer location or perhaps smaller amounts would be lost near the MI608 transfer location downstream.

A series of steps was applied successively to mitigate these problems. The Main Injector damping system[13] was applied in anti-damping mode to the injection buckets as permitted by the phases of the slip stacking process. This was helpful but an incomplete solution. In 2010, kicker magnets ware installed[14] to allow this gap to be cleared by transferring this beam to the Main Injector abort one turn prior to the next injection.

For the beam gap for extractions, the anti-damping solution was less constrained. It was applied initially with good success using the regular damping kickers. A magnet installed in the injection region was employed to provide a stronger kick which allowed loss-free operation at the MI522 transfer point for extended periods of time.

## FINDING PROBLEMS

Tuning to optimally employ the collimation, anti-damping and gap clearing kicker improvements described above continued for a period of several years. The Main Injector specialists and the accelerator operation crew employed the loss display and other tools to progressively limit the locations where significant loss occurred. As a result one could note that the major losses occurred early in the cycle and were concentrated in the collimation region and at transfer points. We were now free to examine a limited number of 'unexplained' losses.

As an example, we were concerned by losses near Q113, a vertically focusing regular half cell. Comparisons of the aperture indicated a nearly normal horizontal aperture but a substantially reduced vertical aperture. Due to the power and stability of the orbit correction, the effect of this aperture restriction created small losses and small but significant residual radiation. External mapping of the residual radiation pattern provided subtle hints of some aperture restriction. When time and manpower became available, the beam pipe was opened and a bore scope used to examine the entire region from the last dipole in MI112 to the first dipole in MI113. A small problem was discovered upstream but the major limitation was due to the 'rf fingers' which span the gap created by the formed bellows. During an installation eight years earlier, the pipe had been tipped such that these fingers escaped into the beam aperture, reducing the vertical aperture by about 5 mm. Replacement of the faulty bellows resulted in immediate elimination of the loss signal which had be apparent at the loss monitors at 113,114 and 115.

## RESULTS

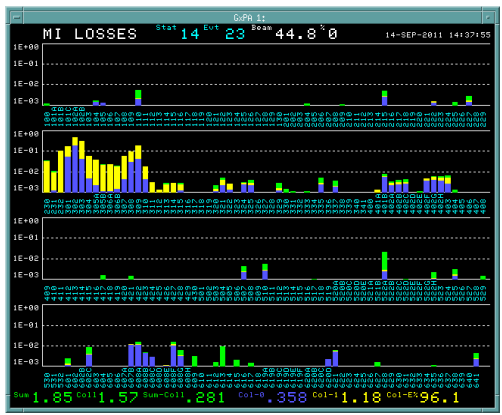

Figure 4: Beam Loss Monitor Display from September 2011 at intensity of $4.5 \times 10^{13}$ protons. Loss lower everywhere except at collimators.

Successes in reducing loss were gratifyingly apparent with the loss display shown in Figures 2 and 4. For the last years of running, we had successfully attacked many losses, eliminating some entirely.

The definitive measure of loss control is reduced radioactivity for hands-on maintenance and upgrade activities. On the 3.3 km scale of the Main Injector, localization of the loss implies that no single measure of radiation reduction will describe the impact of improvements. The successes of the loss control campaign in the Main Injector has lead to enormous improvements in all regions except at the collimators. We will illustrate this in two ways. Figures 5 and 6 provide snapshots of the residual radiation at bar-coded locations. Note the logarithmic scale where a reduction by a factor of ten shows with a reduction of a bar by 1/3 of the vertical scale. We see, as we did with the loss display, the residual radiation is greatly reduced. Only the collimator region remains at a nearly constant residual rate. Beware of detailed comparisons since the measurements are done with various delays after the end of beam operations.

For each bar coded location, we have a series of more than 40 measurements from 2005 through 2012. Control console programs allow one to display these results and to fit them to the half-life-weighted BLM readings available for each beam pulse[15].Figure7 shows a most atypical such plot for the residual radiation at the extraction point for antiproton production. High losses associated with commissioning of slip stacking and NuMI start-up were mitigated with collimation removal of kicker gap beam such that loss free operation was achieved for extended periods of time. Much of the remaining residual radiation is still from activation many years ago. For more typical locations, the radiation reduction during multiweek shutdowns are apparent. Tools to show integrated loss (see HB2010 presentation[15]) allow examination of any BLM to allow comparison of losses with any observation of new high radiation points. These tools were used for radiation exposure planning for the 2012 facility shutdown. Active monitoring of residual radiation continues.

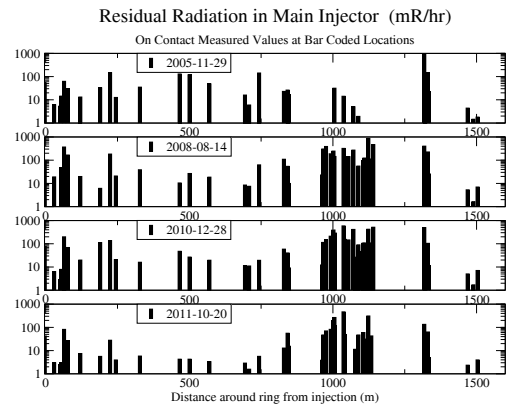

Figure 5: Residual Radiation History from Injection to Abort Region.

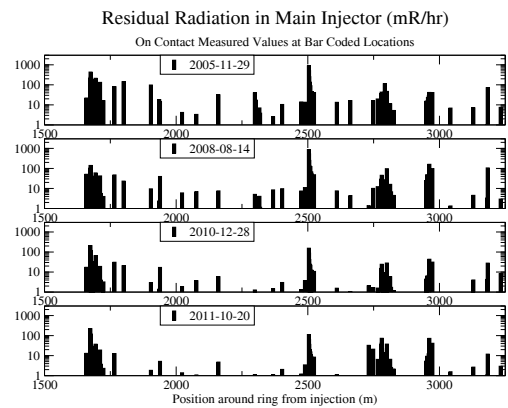

Figure 6: Residual Radiation History from Abort Region to Injection.

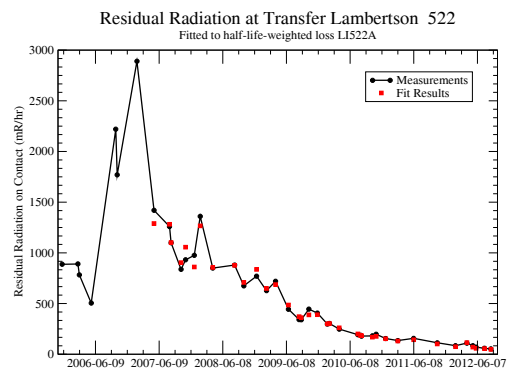

Figure 7: Residual Radiation History at Upstream End of Lambertson LAM522 (transfer and extraction).

## CONCLUSIONS

A campaign of loss control for 400 kW operation of the Fermilab Main Injector has been successfully carried out. Major pieces of this effort include the wide aperture quadrupole upgrade, the collimation efforts and the antidamping and gap clearing kicker system to localize the most significant losses. Instrumentation upgrades allowed observation and mitigation of additional losses. Operational diligence by Accelerator Operations and Main Injector Department staff achieved a loss pattern with few unexpected losses. These locations were then addressed one-by-one.

A few observations are appropriate:

- Understanding of loss mechanisms through measurement and simulation was essential.
- The large effort required for major items was well rewarded.
- Finding simple solutions for widespread minor problems allowed one to 'clear the field' and see the rest of the issues.
- Most but not all minor problems were understood. Some problems remain. Some problems disappeared, perhaps because of the success of the more general solutions.

Finally, I comment that clean living is worthwhile.

The Fermilab accelerators entered a shutdown for upgrades on April 30, 2012. The new operation will employ slip stacking in the permanent magnet Recycler Ring with recapture using high rf voltage in the Main Injector. This will permit Main Injector operation at 1.3 second cycle time compared with a 2.2 second cycle time. New locations and perhaps new loss mechanisms are expected. The uncaptured beam loss will be localized with the same Main Injector collimation system. The tools we have developed are expected to provide a sound way forward to allow the design operation at 700 kW of proton beam power to be achieved without unmanageable activation.